\documentclass[aps,amsmath,amssymb,reprint,superscriptaddress]{revtex4-2}

\usepackage{graphicx}
\usepackage{dcolumn}
\usepackage{bm}
\usepackage{xcolor}

\usepackage{soul}
\newcommand{\nocontentsline}[3]{}
\let\origcontentsline\addcontentsline
\newcommand\stoptoc{\let\addcontentsline\nocontentsline}
\newcommand\resumetoc{\let\addcontentsline\origcontentsline}

\begin{document}

\preprint{APS/123-QED}

\title{Strong Intrinsic Longitudinal Coupling in Circuit Quantum Electrodynamics}

\author{C.A. Potts}
\email{clinton.potts@nbi.ku.dk}
\affiliation{Kavli Institute of Nanoscience, Delft University of Technology, PO Box 5046, 2600 GA Delft, The Netherlands}
\affiliation{Niels Bohr Institute, University of Copenhagen, Blegdamsvej 17, 2100 Copenhagen, Denmark}
\affiliation{NNF Quantum Computing Programme, Niels Bohr Institute, University of Copenhagen, Denmark}

\author{R.C. Dekker}
\affiliation{Kavli Institute of Nanoscience, Delft University of Technology, PO Box 5046, 2600 GA Delft, The Netherlands}

\author{S. Deve}
\affiliation{Kavli Institute of Nanoscience, Delft University of Technology, PO Box 5046, 2600 GA Delft, The Netherlands}

\author{E.W. Strijbis}
\affiliation{Kavli Institute of Nanoscience, Delft University of Technology, PO Box 5046, 2600 GA Delft, The Netherlands}

\author{G.A. Steele}
\email{g.a.steele@tudelft.nl}
\affiliation{Kavli Institute of Nanoscience, Delft University of Technology, PO Box 5046, 2600 GA Delft, The Netherlands}

\date{\today}

\begin{abstract}
Radiation-pressure interactions between harmonic oscillators have enabled exquisite measurement precision and control, made possible by using strong sideband drives, enhancing the coupling rate while also linearizing the interaction. In this letter, we demonstrate a strong intrinsic longitudinal coupling, a circuit quantum electrodynamics analogue of the radiation-pressure interaction, between a transmon qubit and a linear microwave resonator. A red-detuned sideband drive results in an on-demand Jaynes-Cummings interaction with a high on-off ratio. We measure a longitudinal coupling rate an order of magnitude larger than all decay rates, placing the device in the strong coupling regime. The intrinsic longitudinal interaction demonstrated here will enable the development of high-connectivity quantum information processing hardware and the exploration of the gravitational decoherence of quantum objects.
\end{abstract}

\maketitle

\textit{Introduction} --- Radiation-pressure interactions arise when the frequency of a system is modulated by the coordinate of a second system. Interactions of this type between harmonic oscillators have provided the foundation for a wide range of groundbreaking experiments in the last several decades. A canonical example is that of cavity optomechanics \cite{aspelmeyer2014cavity} but includes related fields such as cavity magnomechanics \cite{zhang2016cavity, potts2021dynamical, shen2022mechanical, potts2023dynamical}, photon-pressure coupling \cite{eichler2018realizing, bothner2021photon, rodrigues2021cooling, rodrigues2022parametrically, rodrigues2022photon}, and the center of mass motion of trapped ions \cite{leibfried2003experimental,haljan2005spin}. Such interactions have enabled scientific achievements such as ground-state cooling of mechanical objects \cite{seis2022ground, teufel2011sideband, noguchi2016ground}, the generation of deterministic mechanical entanglement \cite{kotler2021direct}, microwave-to-optical frequency conversion \cite{andrews2014bidirectional, jiang2020efficient}, quantum non-demolition measurements \cite{hertzberg2010back, lecocq2015quantum}. These demonstrations have relied on enhancing the typically low coupling rate via strong sideband drive tones, linearizing the intrinsically nonlinear interaction. 

In the context of superconducting qubits, the interaction analogous to radiation-pressure is known as longitudinal coupling, and has recently been proposed as a new paradigm for quantum information processing \cite{billangeon2015circuit, Johansson2014Optomechanical} with exciting potential applications for quantum processors \cite{Richer2016Circuit,wang2019ideal}, and has been widely demonstrated in ion-trap hardware \cite{leibfried2003experimental, haljan2005spin,sorensen2000Entenglement}. For example, longitudinal coupling could allow for all-to-all qubit connectivity and pure quantum non-demolition qubit readout \cite{didier2015fast, chapple2024robustness, chessari2024unifying}. The application of coherent drives has enabled an effective longitudinal coupling \cite{eickbusch2022fast}, which has been used to demonstrate efficient qubit readout \cite{ikonen2019qubit,touzard2019gated}. However, despite this interest, limited experimental progress on pure longitudinal coupling in superconducting circuits has been made, with most experimental realizations suffering from dominant parasitic transverse coupling spoiling the pure longitudinal nature \cite{chiorescu2004coherent}.

In this article, we demonstrate an entirely hardware-realized  \textit{pure longitudinal} interaction between a superconducting qubit and a linear microwave resonator circuit. The device implements longitudinal coupling with negligible transverse coupling encoded directly in the circuit design. The interaction between the qubit and the resonator is engineered by threading the magnetic flux generated by the linear resonator through a superconducting quantum interference device (SQUID) \cite{eichler2018realizing,bothner2021photon, Johansson2014Optomechanical}, forming the inductance of the qubit. 
Using a sideband drive, we induce an effective Jaynes-Cummings interaction in the strong coupling limit which can be dynamically controlled with a high on-off ratio. Furthermore, the AC-Stark shift of the qubit provides an absolute photon number calibration \cite{Ann_Sideband_2021}, allowing for the extraction of the longitudinal coupling rate. We find a longitudinal coupling rate, $g_0 = 2\pi \times 11.9$ MHz, two orders of magnitude larger than the largest dissipation rate of the system, placing our system deep in the single-photon strong coupling regime \cite{aspelmeyer2014cavity}.

\begin{figure*}[t!]
    \centering
    \includegraphics[width = 0.95\textwidth]{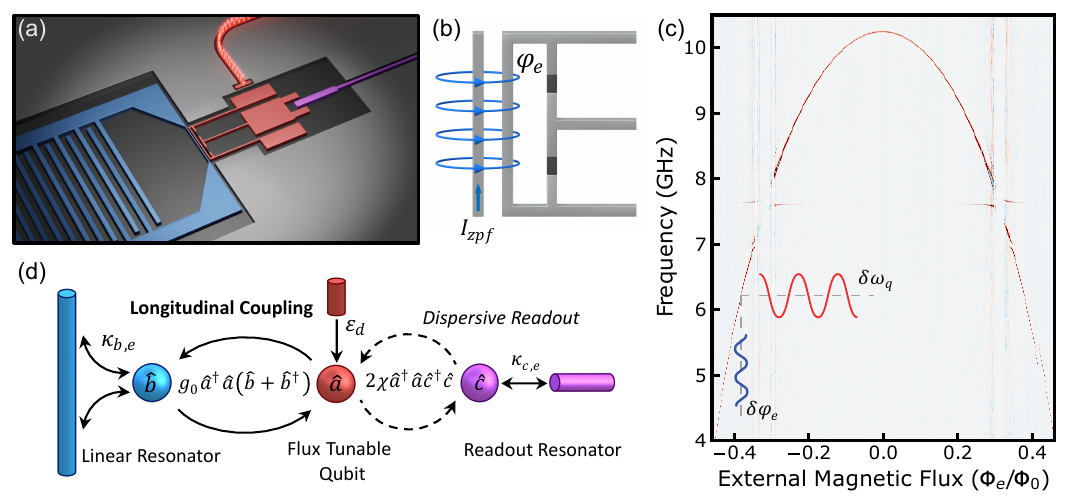}
    \caption{A superconducting quantum circuit with pure longitudinal coupling. (a) A rendering of the superconducting device. (b) Schematic depiction of the longitudinal coupling mechanism. Zero-point current in the linear resonator generates a magnetic flux that modulates the qubit frequency. (c) Qubit two-tone spectroscopy as a function of the external magnetic flux. The qubit has a sweet spot frequency of $10.2$ GHz. The magnetic flux generated by currents in the linear resonator modulates the qubit frequency, resulting in longitudinal coupling. Data near the readout resonator frequency of $7.6$ GHz is excluded due to the qubit resonator hybridization. (d) Schematic diagram of the device. The qubit $\hat{a}$ is coupled via a longitudinal interaction to the linear resonator $\hat{b}$ and coupled capacitively to a CPW resonator $\hat{c}$.}
    \label{Figure1}
\end{figure*}

The single-photon strong coupling regime has long been sought as it fully utilizes the inherent nonlinearity of the radiation-pressure interaction \cite{nunnenkamp2011single,nunnenkamp2012cooling}. Since linear interactions can only perform bilinear operations on Gaussian states, single-photon strong coupling would allow the dissipative engineering of states with Wigner negativity, such as described in the protocol for the generation of Schr\"{o}dinger cat-states \cite{hauer2023nonlinear}. Moreover, the longitudinal interaction allows a large number of photons to be driven into the linear resonator, unlike in conventional transverse cQED \cite{Krantz_Quantum_2019}. Therefore, combining longitudinal cQED with conventional optomechanics integrated within the linear resonator will enable the generation of mechanical Schr\"{o}dinger cat states, via a double state swap procedure, to explore gravity's effect on quantum mechanics \cite{gely2021superconducting}.

\textit{Device Design} --- The device consists of a modified transmon qubit coupled to two superconducting resonators, all with frequencies in the gigahertz range. The device's full circuit schematic can be seen in Fig.~\ref{Figure1}. All circuits are fabricated from niobium titanium nitride \cite{thoen2016superconducting}; the fabrication details are outlined in the supplementary information  \cite{supp}. A $\lambda/2$ co-planar waveguide resonator (CPW) labelled $\hat{c}$ is capacitively coupled to the qubit and acts as a standard readout resonator for qubit measurements and state preparation \cite{Krantz_Quantum_2019,blais2021circuit}. It has a frequency $\omega_{\rm c} = 2 \pi \times 7.612$ GHz, a qubit-photon coupling rate $g_{\rm ac} = 2\pi \times 118$ MHz, an internal and external coupling rate of $\kappa_{\rm int,c} = 2 \pi \times 72.8$ kHz and $\kappa_{\rm ext,c} = 2 \pi \times 673.8$ kHz.

The linear resonator labelled $\hat{b}$ consists of a large interdigitated capacitor (IDC) shunted with a thin inductor wire. The linear resonator has a resonance frequency $\omega_{\rm b} = 2 \pi \times 4.347$ GHz, an internal and external coupling rate of $\kappa_{\rm int,b} = 2 \pi \times 28.0$ kHz and $\kappa_{\rm ext,b} = 2 \pi \times 88.6$ kHz and a residual parasitic dipolar coupling to the qubit $g_{\rm ab} = 2 \pi \times 1.4$ MHz. The parasitic capacitive coupling was minimized by careful qubit design and results in a critical photon number at the qubit operational point of $n_{\rm crit} \equiv \Delta^2 / 4 g_{\rm ab}^2 \approx 350\,000$, where $\Delta$ is the qubit-linear resonator detuning \cite{Krantz_Quantum_2019}. The reduced hybridization between the qubit and linear resonator is an essential feature of the qubit design; the low coupling rate $g_{\rm ab}$ allows relatively large coherent drives to be applied to the linear resonator. We observe a bare-to-dressed transition of the linear resonator with a steady state drive $\vert \langle \hat{b} \rangle \vert^2 \approx 20\,000$, which is approximately an order of magnitude smaller than the critical photon number. However, we expect the number of drive photons to increase significantly even with moderate reductions of the parasitic dipolar coupling \cite{boissonneault2009dispersive}. These large coherent drives will be necessary if the linear resonator is replaced with a cavity optomechanical device since large coherent tones are required for mechanical state manipulation \cite{teufel2011sideband}.

Finally, the qubit labelled $\hat{a}$ is a modified pocket-style transmon qubit with Manhattan-style aluminum/aluminum-oxide Josephson junctions \cite{Qiskit}. As seen in Fig.~\ref{Figure1}, one of the capacitor pads of the transmon is split and wrapped around the other, and the SQUID loop is extended away from the capacitor pads. The SQUID loop is placed $\sim 1 \mu$m from the thin inductor wire of the linear circuit. This design was implemented to reduce the electric dipole moment of the qubit, minimizing the parasitic dipole coupling to the linear resonator, as discussed previously. We believe the parasitic coupling likely results from junction asymmetry resulting in finite current through the linear inductor of the SQUID and charging of the surrounding ground plane. 

A coil mounted below the sample provides an external magnetic field that can tune the frequency of the qubit; see Fig.~\ref{Figure1}(c). During this work, the qubit was operated at a frequency $\omega_{\rm q}(\Phi_{\rm e}) \sim 2\pi \times 6.10$ GHz, where $\Phi_{\rm e}$ is the static external magnetic flux thread through the SQUID. The qubit can be driven via a direct XY-drive line, which was designed to be weakly coupled such that the Purcell limited linewidth $\kappa_{\rm pur}$ is much less than the qubit linewidth $\gamma_{\rm q}$. Finally, we measured a $T_2^*$ limited linewidth $\gamma_{\rm q} = 2 \pi \times 677$ kHz; qubit characterization is described within the supplementary information. The chip is mounted on the baseplate of a commercial dilution refrigerator operating with a base temperature of $T \sim 15$ mK. For details about the measurement and set-up, see the supplementary information  \cite{supp}. 

\textit{Operational Principle} --- We model the qubit as a Kerr oscillator with a large anharmonicity \cite{Krantz_Quantum_2019}, the bare Hamiltonian of the system can be written in the form,
\begin{equation}
    \mathcal{\hat{H}}_0/\hbar = \omega_{\rm q}(\Phi_{\rm e})\hat{a}^{\dagger}\hat{a} + \frac{\alpha}{2}\hat{a}^{\dagger}\hat{a}^{\dagger}\hat{a}\hat{a} + \omega_{\rm b} \hat{b}^{\dagger}\hat{b}.
\end{equation}
Here, $\alpha = -388$ MHz is the qubit anharmonicity at the operation point, $\hat{a}^{(\dagger)}$ and $\hat{b}^{(\dagger)}$ are the qubit and linear resonator annihilation (creation) operators, respectively.

Current flowing through the inductor wire of the linear resonator generates an oscillating magnetic flux through the SQUID loop, modulating the qubit's frequency and inducing the coupling between the qubit and linear resonator, see Fig.~\ref{Figure1}(b,c). It can be shown, after some algebraic manipulation and keeping only first-order terms --- see the supplementary information --- that the interaction can be written in the simple form \cite{supp},
\begin{equation}
    \mathcal{\hat{H}}_{\rm int}/\hbar = g_0\hat{a}^{\dagger}\hat{a}(\hat{b} + \hat{b}^{\dagger}).
    \label{Non_lin_Int}
\end{equation}
Here, $g_0$ is the longitudinal coupling rate and is given by
\begin{equation}
    g_0 = \frac{\partial \omega_{\rm q}}{\partial \Phi}\Phi_{\rm zpf},
\end{equation}
where $\partial\omega_{\rm q}/\partial\Phi$ is the qubit's flux sensitivity at the operational point, and $\Phi_{\rm zpf}$ is the magnetic flux thread through the SQUID loop generated by the zero-point current fluctuations of the linear resonator. The zero-point flux can be estimated from the mutual inductance and the zero-point fluctuations in the current given by $I_{\rm zpf} = \sqrt{\hbar \omega_{\rm b}/2L_{\rm b}} \approx36.8$ nA; with $\Phi_{\rm zpf} = M I_{\rm zpf}$. The mutual inductance can be estimated via the Biot-Savart law and gives $M \approx 25.9$ pH; therefore, we estimate the zero-point magnetic flux to have a value $\Phi_{\rm zpf} \approx 461 \, \mu\Phi_{0}$. We have normalized the magnetic flux in units of $\Phi_0 = h / (2e)$. At the qubit operational point, the flux sensitivity $\partial\omega_{\rm q}/\partial\Phi = 2 \pi \times 26.0$ GHz$/ \Phi_0$, thus we can estimate the longitudinal coupling rate to have a value $g_0 \approx 2 \pi \times 12.0$ MHz. 

\begin{figure}[b]
    \centering
    \includegraphics[width = 0.45\textwidth]{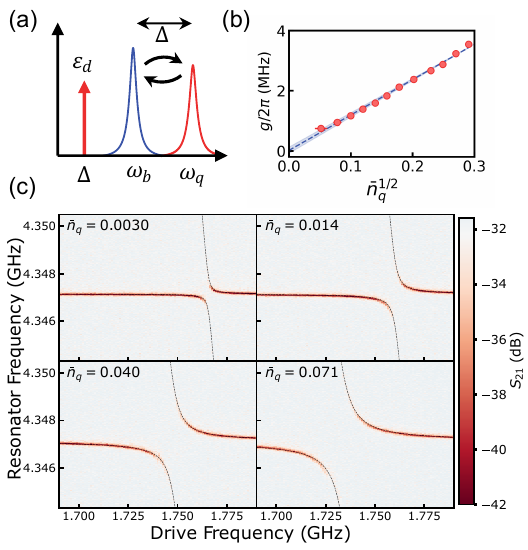}
    \caption{On-demand strong coupling by driving the longitudinal interaction. (a) Schematic diagram of the relative frequencies of the qubit (red), the linear resonator (blue), and the sideband drive (red). (b) Cavity-enhanced coupling rates extracted from the normal mode spectra at different sideband drive powers. The coupling rates are plotted as a function of the square root sideband photon number. The red points are data, and the blue dashed line is a linear fit to the data. The blue shaded area accounts for uncertainty in the bare qubit frequency. The photon number was calibrated using the AC-Stark shift. (c) The measured avoided level crossings of the linear resonator as a function of the sideband drive frequency. Starting from the top left, the drive powers considered here were $P_{\rm sb} = 3.3, 8.3, 12.3, 14.4$ dBm, set at the source. The calibrated average number of sideband photons is printed in the top left of each panel. The dashed line is a fit; see the supplementary information for details.}
    \label{Figure3}
\end{figure}

\begin{figure*}[t]
    \centering
    \includegraphics[width = 0.95\textwidth]{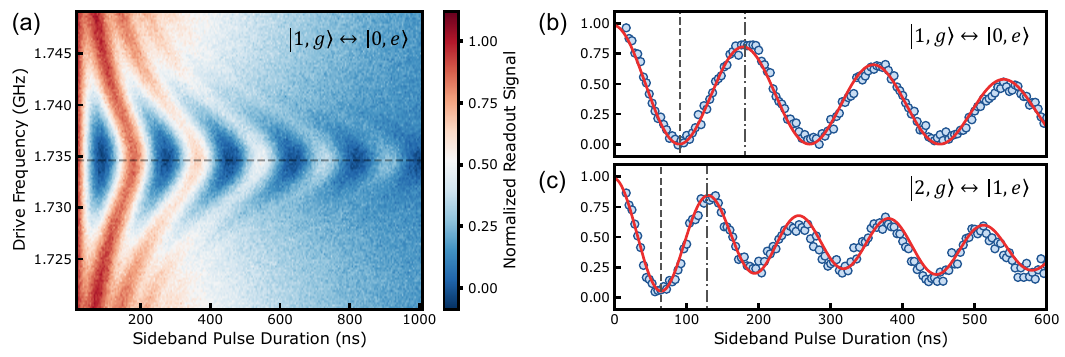}
    \caption{Observation of coherent swaps of photons up the on-demand Jaynes Cummings ladder (a) Rabi chevrons with a single excitation in the system. The normalized output signal is proportional to the excited state probability of the qubit $P_{\vert e \rangle}$. (b) Linecut of resonant Rabi oscillations for the $\vert 1,g \rangle \leftrightarrow \vert 0,e \rangle$ transition. A Rabi frequency $g = 2\pi \times 2.76$ MHz was used in the numerical simulation. The dashed lines indicate the longitudinal pi-pulse duration $t_{\pi} = 90.5$ ns. The dotted dashed lines indicate the Rabi oscillation period, and the solid red line is a numerical simulation of the Rabi oscillations. (c) Linecut of resonant Rabi oscillations for the $\vert 2,g \rangle \leftrightarrow \vert 1,e \rangle$ transition. An increased Rabi frequency $g_2 \approx \sqrt{2} g$ is observed. The dashed lines indicate the predicted second manifold longitudinal pi-pulse duration $t_{\pi,2} = 90.5 / \sqrt{2}$ ns. The dotted dashed lines indicate the Rabi oscillation period, and the solid red line is a numerical simulation of the Rabi oscillations.} 
    \label{Figure5}
\end{figure*}

\textit{On-Demand Jaynes-Cummings} --- The on-demand drive-induced Jaynes-Cummings interaction was implemented using a sideband drive tone tuned to the difference frequency between the two coupled oscillators; see Fig.~\ref{Figure3}(a). In cavity optomechanics, the sideband drive is typically nearly resonant with the optical cavity, as mechanical frequencies are 1-10 MHz for most electromechanical systems \cite{seis2022ground,noguchi2016ground}. However, this device operates in a vastly different parameter regime. The sideband drive frequency is below both the qubit and linear resonator frequencies. We should emphasize that we are driving single-photon sideband transitions and \textit{not} the two-photon sideband transitions used in conventional cQED \cite{wallraff2007sideband}.

We can linearize the Hamiltonian about the sideband drive tone with frequency $\omega_{\rm d} \approx \omega_{\rm q} - \omega_{\rm b}$ tuned near the red sideband applied via the XY-drive line to the qubit. In a displaced frame co-rotating with the drive, we can linearize by defining $\hat{a} = \langle \hat{a} \rangle + \delta\hat{a}$; doing so, we can define the enhanced coupling rate as $g = g_0 \vert \langle \hat{a} \rangle \vert$, and $\Bar{n}_{\rm q} = \vert \langle \hat{a} \rangle \vert^2$ is the average sideband steady-state population of the qubit. The interaction Hamiltonian given in Eq.~\ref{Non_lin_Int} can be written, taking the two-level approximation of the transmon and applying the rotating wave approximation, as
\begin{equation}
    \mathcal{\hat{H}}_{\rm int}/\hbar = g(\hat{\sigma}_-\hat{b}^{\dagger} + \hat{\sigma}_+\hat{b}),
\end{equation}
where $\hat{\sigma}_{\pm}$ are the qubit raising and lowering operators. The linearization process results in an on-demand drive-induced Jaynes-Cummings interaction created by the sideband drive; see the supplementary information for a full derivation  \cite{supp}.

The interaction is experimentally investigated by sweeping a sideband drive near the red sideband and probing the linear resonator with less than a single photon on average to remain in the single-excitation manifold of the Jaynes-Cummings ladder \cite{fink2008climbing}. We observe a series of avoided level crossings with a correspondingly increasing coupling rate as a function of increasing sideband power and a shift to lower frequencies, as shown in Fig.~\ref{Figure3}(c). We can extract the cavity-enhanced coupling rate $g$ from the avoided level crossings. However, we require an absolute calibration of the photon number to determine the longitudinal coupling rate. Conventionally, this is achieved by varying the temperature of the dilution refrigerator and performing thermal calibration \cite{seis2022ground}, or applying a known modulation and carefully measuring all loss and gain within the system \cite{gorodetksy2010determination,macdonald2016optomechanics}. 

Instead, we can use the AC-Stark shift of the qubit for direct photon number calibration. As we increase the sideband drive power, the qubit experiences an increasing AC-Stark shift. The magnitude of the Stark shift is directly related to the average number of sideband photons in the qubit. One must take care when interpreting the Stark shift, which is why we initially modelled the qubit as a Kerr oscillator. Since the sideband drive is far detuned, the counter-rotating terms, known as the Bloch-Siegert shift \cite{Ann_Sideband_2021}, and the qubit anharmonicity contribute significantly. Considering the steady state of the Kerr oscillator and the AC-Stark shift, the steady-state qubit population for each sideband power can be directly calculated.

The value of the longitudinal coupling rate can be determined by considering the cavity-enhanced coupling rate as a function of the square root of the number of steady-state sideband photons, as seen in Fig.~\ref{Figure3}(b). Importantly, the relationship is linear, suggesting our approximations are valid and hold even for steady-state populations below the single photon level, and the data has a y-intercept of zero, which is expected since the cavity-enhanced coupling should vanish for no sideband drive. The value of the longitudinal coupling rate can be determined from the slope, giving a value of $g_0 = 2 \pi \times 11.9 \pm 0.3$ MHz, which is in excellent agreement with our theoretical prediction and well within the single-photon strong coupling regime. 

\textit{Rabi Oscillations} --- Finally, to demonstrate the coherent nature of the interaction, we measure vacuum Rabi oscillations between the qubit and the linear resonator. The qubit is first prepared into its excited state $\vert e \rangle$ using a 24 ns resonant $\pi$-pulse. A square wave sideband pulse is applied near the red sideband frequency for a variable time $\tau$. Following the sideband pulse, the qubit state is measured using the capacitively coupled CPW resonator. This protocol is repeated for various sideband drive frequencies and pulse durations, resulting in the vacuum Rabi oscillations shown in Fig.~\ref{Figure5}. We observe a typical Rabi ``chevron'' pattern with coherent exchanges between the qubit and the linear resonator. A line cut is fit using a numerical simulation of the Rabi oscillations to confirm our linear model; see the supplementary information  \cite{supp}. From the numerical simulation, we can extract the coupling rate $g/2\pi = 2.76 \pm 0.03$ MHz, \cite{hofheinz2008generation} resulting in a sideband $\pi$-pulse duration of $\sim$ 91 ns; see Fig.~\ref{Figure5}(b). This is in excellent agreement with the measured normal mode splitting at nominally the same sideband power, $g/2\pi = 2.81 \pm 0.08$ MHz. Finally, we observe transitions to higher excited manifolds of the Jaynes-Cummings ladder by preparing the linear oscillator in the Fock state $\vert 1 \rangle$ using a sideband $\pi$-pulse and measuring the Rabi oscillations. We observe a Rabi frequency with a value $g_2 \approx \sqrt{2}g$, as predicted by the Jaynes-Cummings model \cite{hofheinz2008generation}; see Fig.~\ref{Figure5}(c). 

\textit{Outlook} --- This work presents a unique interface between a superconducting qubit and a linear microwave resonator. The interaction was confirmed by performing normal-mode spectroscopy of the hybrid modes induced via a red sideband drive and observing coherent Rabi oscillations. The value of the longitudinal coupling rate was determined by calibrating the sideband drive using the AC-Stark shift. Our device is well within the single-photon strong coupling limit.

The innovative qubit design provides a new platform for testing theories demanding single-photon strong coupling, such as dissipative Schr\"{o}dinger cat state generation within the microwave resonator \cite{hauer2023nonlinear}. Moreover, the large critical photon number, $n_{\rm crit} \approx 350\,000$, and the ability to dynamically modulate the coupling should allow for future integration with optomechanics experiments. In such an experiment, the cat state generated within the linear resonator can be swapped into a mechanical oscillator using a strong optomechanical sideband pulse, generating a macroscopic superposition allowing for exotic tests of gravitational decoherence \cite{gely2021superconducting}. Such a swap to the mechanical resonator would \textit{not} be possible using conventional Jaynes-Cummings coupling due to the large driving requirements on the linear microwave cavity.

Moreover, this platform may also be used for interesting on-chip quantum optics experiments; for example, at the top sweet-spot to first order, the linear coupling should be suppressed, \textit{i.e.} $g_0 \approx 0$; however, there will remain quadratic coupling of the form $\mathcal{\hat{H}}_{\rm q}/\hbar = g_{\rm q}\hat{a}^{\dagger}\hat{a}(\hat{b}+ \hat{b}^{\dagger})^2$. Quadratic coupling has been long sought and was an initial driving force for the membrane in the middle experiments \cite{nunnenkamp2010cooling}. By eliminating linear coupling, the quadratic coupling may allow quantum non-demolition measurement of the photon number within the linear resonator. This enables an on-chip cQED measurement of stochastic photon jumps as they decay from the linear resonator \cite{guerlin2007progressive,gleyzes2007quantum}.

Finally, the longitudinal coupling can be explored as a unique architecture for quantum information systems \cite{Richer2016Circuit,didier2015fast, wang2019ideal,billangeon2015circuit}. Multiple qubits could be coupled to a single bus resonator, and by applying appropriate sideband pulses, sideband drive-mediated qubit-qubit gates may be possible. All qubits connected via the longitudinal interaction to a single bus resonator would allow a fully superconducting implementation of efficient quantum computation architectures currently studied using ion traps \cite{wright2019benchmarking}.

\stoptoc

\begin{acknowledgments}
The authors thank V.A.S.V Bittencourt and M. Kjaergaard for helpful discussions and Enrique Sahagun for the device rendering \cite{Scixel}. C.A.P. acknowledges the support of the Natural Sciences and Engineering Research Council of Canada (NSERC) (PDF-567689-2022) and the Novo Nordisk Foundation, NNF Quantum Computing Programme. R.C.D. acknowledges support from the Netherlands Organisation for Scientific Research (NWO/OCW) as part of the Frontiers of Nanoscience program.

This publication is part of the project `Superconducting Electromechanics: Massive superpositions for exploring quantum mechanics and general relativity.' project number VI.C.212.087 of the research programme VICI round 2021, financed by the Dutch Research Council (NWO).

\textbf{Authors contributions:} C.A.P. performed theoretical modeling, performed data analysis, designed the qubit, performed daily supervision of measurements and coordination of the team, wrote the first draft of the manuscript with input from all authors, and incorporated feedback from authors into the final manuscript. R.C.D. designed and simulated the full device layout, developed the fabrication recipe and fabricated the device, performed spectroscopic measurements, contributed to data analysis, and contributed to figure design. S.D. contributed to the verification of theoretical models. S.D. and E.S. performed time-domain qubit measurements and analysis of those measurements. G.A.S. was responsible for conceiving the experiment, overall supervision of the project, and the acquisition of funding for the project. All authors contributed to the formulation of the manuscript storyline and the composition of the figures.

\textbf{Competing interests:} The authors declare no competing interests. 

\textbf{Data and materials availability:} All data, analysis code, and measurement software are available in the manuscript or the supplementary information or are available at Zenodo \url{https://doi.org/10.5281/zenodo.14501323} \cite{Zenodo}.

\end{acknowledgments}
\clearpage

\bibliography{apssamp}

\resumetoc
\clearpage
\onecolumngrid

\setcounter{equation}{0}
\setcounter{figure}{0}
\setcounter{table}{0}
\renewcommand{\theequation}{S\arabic{equation}}
\renewcommand{\thefigure}{S\arabic{figure}}

\noindent\textbf{\textsf{\Large Supplementary Information: Strong Intrinsic Longitudinal Coupling in Circuit Quantum Electrodynamics}}

\vspace{2mm}

\normalsize
\vspace{.3cm}

\noindent\textsf{C.A.~Potts$^{1,2,3}$, R.C.~Dekker$^1$,  S.~Deve$^1$, E.W.~Strijbis$^1$, G.A.~Steele$^1$}

\vspace{.2cm}
\noindent$^1$\textit{Kavli Institute of Nanoscience, Delft University of Technology, PO Box 5046, 2600 GA Delft, The Netherlands}\\
\noindent$^2$\textit{Niels Bohr Institute, University of Copenhagen, Blegdamsvej 17, 2100 Copenhagen, Denmark}\\
\noindent$^2$\textit{NNF Quantum Computing Programme, Niels Bohr Institute, University of Copenhagen, Denmark}\\

\addtocontents{toc}{\protect\setcounter{tocdepth}{0}}
\tableofcontents

\section{Materials and Methods}

\subsection{Device Fabrication}

The device is fabricated on 10x10 mm 525 $\mu$m thick high-resistivity silicon chips with 200 nm of niobium titanium nitride (NbTiN). The NbTiN is deposited by the Dutch Institute for Space Research (SRON) following the procedure described in Ref.~\cite{thoen2016superconducting}. The bulk of the circuitry is fabricated by spinning a (positive) e-beam resist layer (AR-P 6200.18, 4000 rpm) and patterning the CPW, linear resonator, qubit islands and feedlines using electron-beam lithography. After development (Pentylacetate, O-xylene, IPA) and a post-exposure bake, the NbTiN is removed using an SF$_6$ reactive ion etching step, followed by an in-situ oxygen plasma descum. After stripping the resist, the Josephson junctions are fabricated by depositing an MMA/PMMA resist bi-layer and patterning the junctions using electron-beam lithography. The resist is developed using cold H\textsubscript{2}O:IPA (1:3) and IPA. The chip is dipped in a buffered oxide etchant (BOE) for 1 min prior to loading into the Plassys deposition system. The junctions are evaporated using a double-angle deposition technique, forming a pair of Manhattan-style Josephson junctions. After liftoff in NMP and wire bonding, the chips are loaded into the dilution refrigerator.

\subsection{Measurement Setup}

\begin{figure*}[ht!]
    \centering
    \includegraphics[width = 1.00\textwidth]{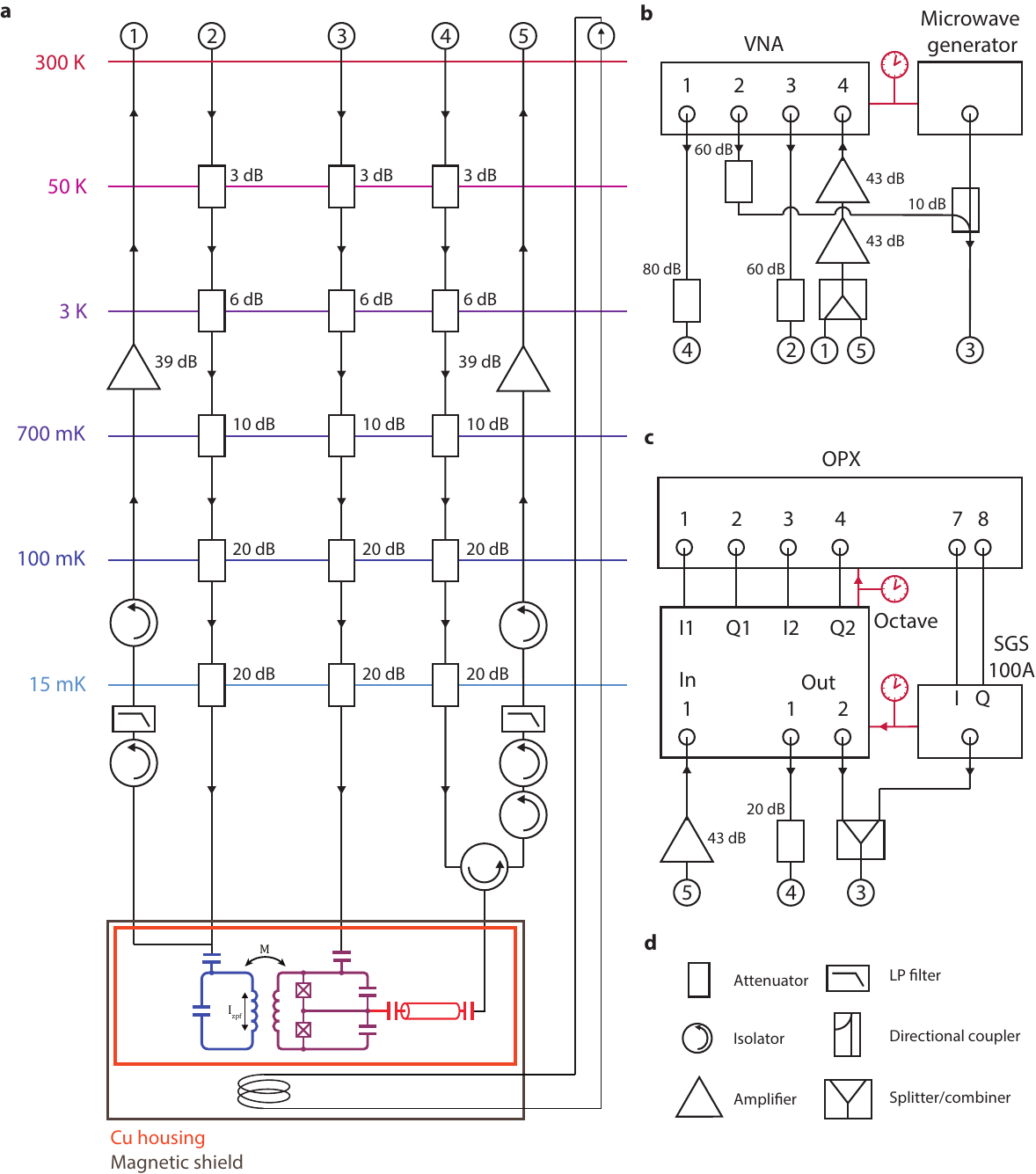}
    \caption{\textbf{Schematic of the measurement setup.} \textbf{a}, Wiring diagram inside the dilution refrigerator. \textbf{b}, Room temperature setup for VNA measurements. \textbf{c}, Setup for time domain measurements. \textbf{d}, Overview of components.}
    \label{SIFig_Setup}
\end{figure*}

\clearpage

All measurements were performed on the base plate of a dilution refrigerator operating at a base temperature $T \sim 15$ mK. A complete schematic of the wiring within the fridge is shown in Fig.~\ref{SIFig_Setup}. The $\lambda$/2 CPW readout resonator is connected to its own feedline via a cryogenic circulator and measured in reflection. The linear resonator is coupled in a notch-type geometry and measured in transmission using a different output line. Both output lines are passed through a pair of low-temperature isolators and via superconducting coaxial cables to a pair of HEMT amplifiers and additional amplification at room temperature. The input lines connecting to the $\lambda$/2 CPW resonator, the qubit drive line, and the input line of the linear resonator all have 59 dB of cryogenic attenuation, with additional room-temperature attenuation. The device is mounted in a light-tight copper box on a copper bracket within a superconducting aluminum and Mu-metal magnetic shield. An external superconducting coil magnet is mounted underneath the copper sample box and inside the magnetic shielding. Spectroscopic measurements were performed using a four-port Keysight PNA for both the qubit and linear resonator, and the time domain measurements were performed using a Quantum Machines OPX, Octave, and Rohde $\&$ Schwarz SGS100A vector microwave generator.

\begin{figure*}[t]
    \centering
    \includegraphics[width = 0.9\textwidth]{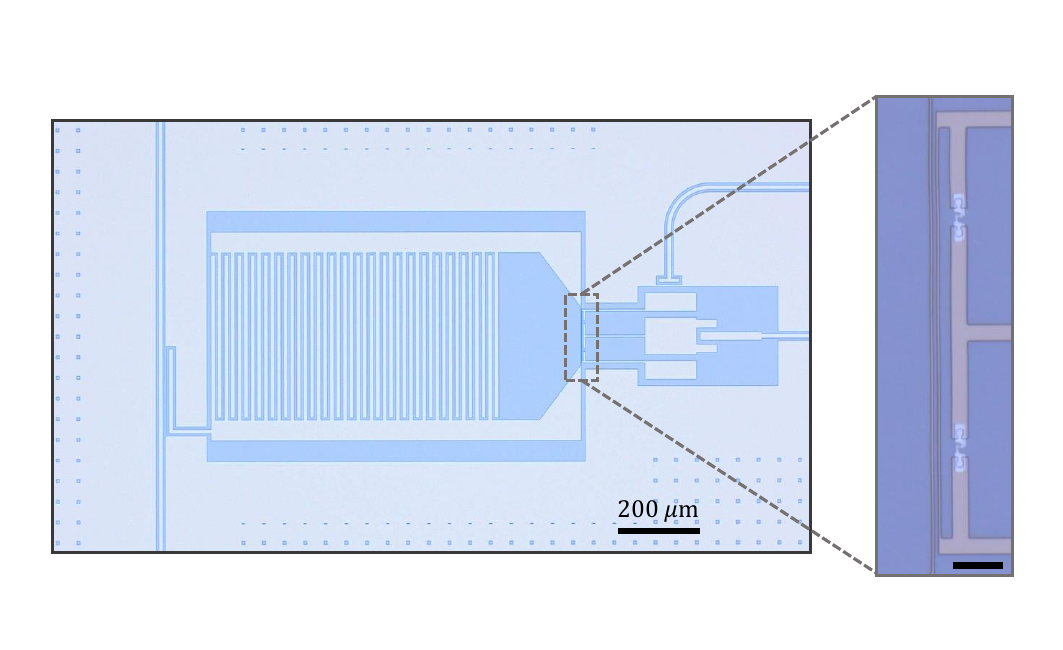}
    \caption{\textbf{Superconducting circuit.} \textbf{a}, Optical micrograph image of the qubit. The scale bar on the zoom is 15 $\mu$m.}
    \label{SIFig_Image}
\end{figure*}

\subsection{Zero-Point Current Fluctuations}

The linear RF circuit consists of a set of interdigitated capacitors (IDC) with a thin inductor wire; see Fig.~\ref{SIFig_Image}. The capacitance of the IDC can be estimated analytically following Ref.~\cite{igreja2004analytical}, 
\begin{equation}
    C_{\rm IDC} = (N-3)\frac{C_1}{2} + 2\frac{C_1C_2}{C_1+C_2}
\end{equation}
where
\begin{equation}
    C_{\rm i} = 2\epsilon_0\epsilon_{\rm eff}l\frac{K(k_{i})}{K(k_{i}')}, \:\:\: i=1,2.
    \label{IDC_Cap}
\end{equation}
Here, $K(k_{i})$ are the elliptic integrals of the first kind, $l$ is the length of the IDC fingers, $\epsilon_{\rm eff} = (\epsilon_{\rm r} +1)/2$ is the effective permittivity where the permittivity of silicon is $\epsilon_{\rm r} = 11.8$, $N$ is the total number of fingers and
\begin{equation}
    k_1 = \sin\bigg( \frac{\pi}{2}\frac{a}{a+b} \bigg),
\end{equation}
\vspace{10pt}
\begin{equation}
    k_2 = 2\frac{\sqrt{a(a+b)}}{2a+b},
\end{equation}
\begin{equation}
    k_{i}'=\sqrt{1-k_i^2},
\end{equation}
where, $a$ is the finger width, and $b$ is the gap between fingers. The fabricated linear resonator has $N = 44$, $a = 10 \mu$m, $b = 6 \mu$m, and $l = 400 \mu$m. Therefore, Eq.~\ref{IDC_Cap} estimates the capacitance of the linear circuit as $C_{\rm IDC} = 1.26$ pF, which is in excellent agreement with the Ansys capacitance simulation calculating a total capacitance of $C_{\rm b} = 1.29$ pF, suggesting there is little stray capacitance from the remainder of the circuit. The measured resonance frequency of the circuit is $\omega_{\rm b} = 2\pi \times 4.347$ GHz. Assuming the IDC capacitance is the only capacitance in the circuit, we can determine the circuit inductance by
\begin{equation}
    \omega_{\rm b} = \frac{1}{\sqrt{L_{\rm b}C_{\rm b}}}.
\end{equation}
giving a total inductance of $L_{\rm b} = 1.06$ nH. Therefore, the current zero-point fluctuations through the inductor of the linear circuit are given by
\begin{equation}
    I_{\rm zpf} = \sqrt{\frac{\hbar\omega_{\rm b}}{2L_{\rm b}}}
\end{equation}
with the values calculated above, the zero-point current $I_{\rm zpf} \approx 36.8$ nA.

\section{Device Characterization}

\subsection{Linear Circuit Parameter Extraction}

\begin{figure*}[t]
    \centering
    \includegraphics[width = 0.9\textwidth]{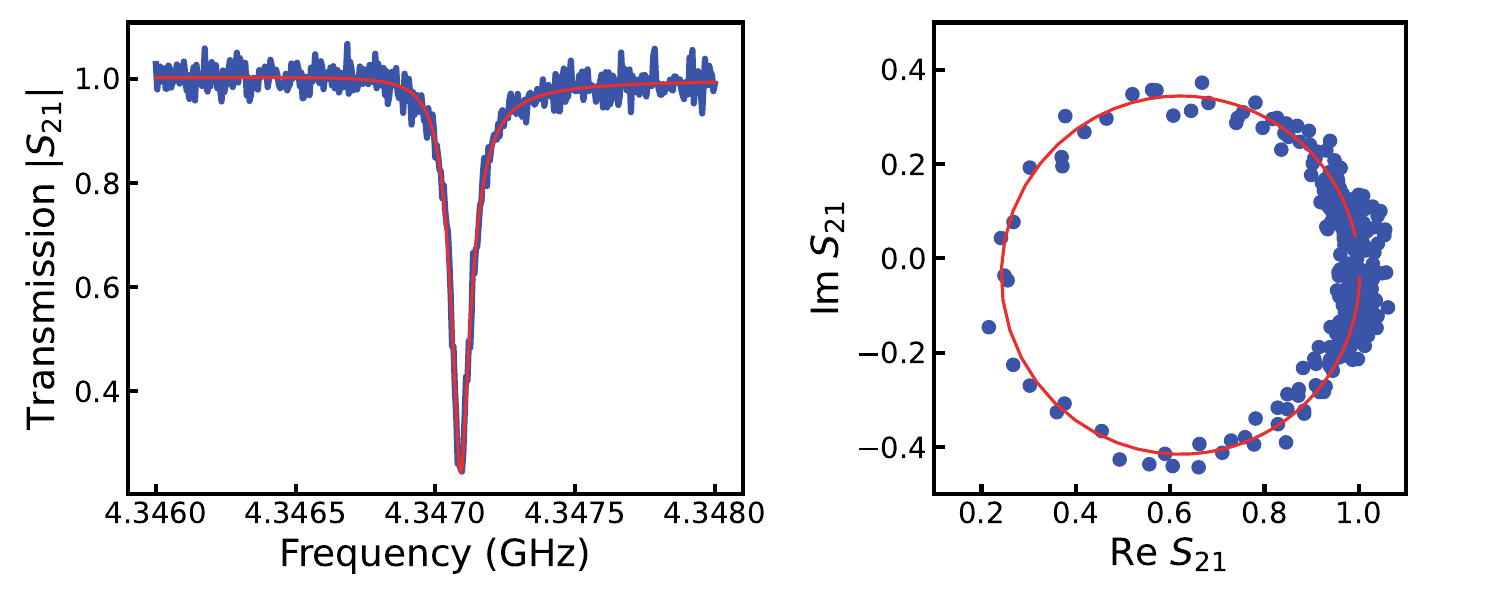}
    \caption{\textbf{The linear circuit response.} \textbf{a}, The magnitude of the transmission response $\vert S_{21}\vert$ of the linear resonator. \textbf{b}, The complex transmission data $S_{21}$ near the resonance frequency of the linear resonator. In both plots, the circles are data; the line is a fit. The data plotted here has had its background subtracted.}
    \label{SIFig_Linear}
\end{figure*}

The linear microwave resonator is coupled to a transmission line in the `notch' style geometry. Thus, the data is fitted to the transmission response function
\begin{equation}
    S_{21}^{\rm notch}(\omega) = 1 - \frac{\kappa_{\rm ext} e^{i\theta}}{(\kappa_{\rm int}+ \kappa_{\rm ext})+2i(\omega - \omega_{\rm b})}.
\end{equation}
Here, $\theta$ accounts for any impedance mismatch, $\omega_{\rm b}$ is the resonance frequency, and $\kappa_{\rm int}$ and $\kappa_{\rm ext}$ are the internal decay rate and the decay rate into the transmission line, respectively. The fit of the measured transmission spectrum is shown in Fig.~\ref{SIFig_Linear}. We extract the parameters $\omega_{\rm b} = 2\pi \times 4.347$ GHz, $\kappa_{\rm int} = 2 \pi \times 28.0$ kHz, $\kappa_{\rm ext} = 2 \pi \times 88.6$ kHz, and $\kappa = 2\pi \times 116.6$ kHz.

\subsection{Qubit Characterization}

The initial characterization of the qubit was done using two-tone spectroscopy. A weak pump tone was applied near the qubit frequency while simultaneously monitoring the frequency of the readout CPW. This technique allows the determination of the qubit frequency and the anharmonicity of the qubit. By sweeping the external magnetic field, we are able to map out the qubit flux arc. We found a sweet spot frequency $\omega_{\rm q} = 2\pi \times 10.2$ GHz and could tune the qubit frequency below $500$ MHz. Next, with the qubit at its operational point, by increasing the qubit pump tone, we are able to measure the anharmonicity by driving the $\vert 1 \rangle \rightarrow \vert 2 \rangle$ transition. We find an anharmonicity $\alpha = -2\pi \times 388$ MHz.

Next, the qubit was characterized by performing time domain measurements of its decay characteristics. At the operational point, the qubit has a $T_1 = 664$ ns and a $T_2^* = 235$ ns. These values are not state-of-the-art; however, they are not a result of the geometry since standard Xmon-style qubits fabricated using the same process show similar values for the characteristic decay rates. We have identified our most likely limiting factor to be our SF$_6$ etch step damaging the silicon surface prior to our junction deposition. This is currently a work in progress, and we aim to improve our coherence time with future devices. Secondly, we expect flux noise to limit our $T_2^*$; this can be improved in future experiments by reducing the SQUID area and optimizing the mutual inductance to maintain a large $g_0$. The current SQUID loop has an area $\sim 360 \mu$m$^2$, likely resulting in large amounts of flux noise. The SQUID loop size can be reduced with device optimization, reducing flux noise. 

\section{Qubit-Resonator Coupling}

\subsection{Parasitic Coupling}
\label{paracoup}

As discussed in the main text, the qubit was designed to minimize dipole coupling between the qubit and the linear resonator. The linear interaction of the form $\mathcal{H_{\rm JC}}/\hbar = g_{\rm ab}(\hat{\sigma}_+ \hat{b} + \hat{\sigma}_-\hat{b}^{\dagger})$ results in the hybridization of the qubit with the linear resonator. Even in the dispersive regime, where $\Delta \gg g_{\rm ab}$, the weak hybridization results in a critical photon number, $n_{\rm crit} = \Delta^2 / (4g_{\rm ab}^2)$, in which the dispersive approximation breaks down \cite{blais2021circuit}. Here, $\hat{\sigma}_{\pm}$ are the qubit raising and lowering operators, $\hat{b}^{(\dagger)}$ are the linear circuit annihilation (creation) operators and $\Delta = \vert \omega_{\rm q} - \omega_{\rm b} \vert$ is the qubit detuning from the linear resonator. 

\begin{figure*}[t]
    \centering
    \includegraphics[width = 0.95\textwidth]{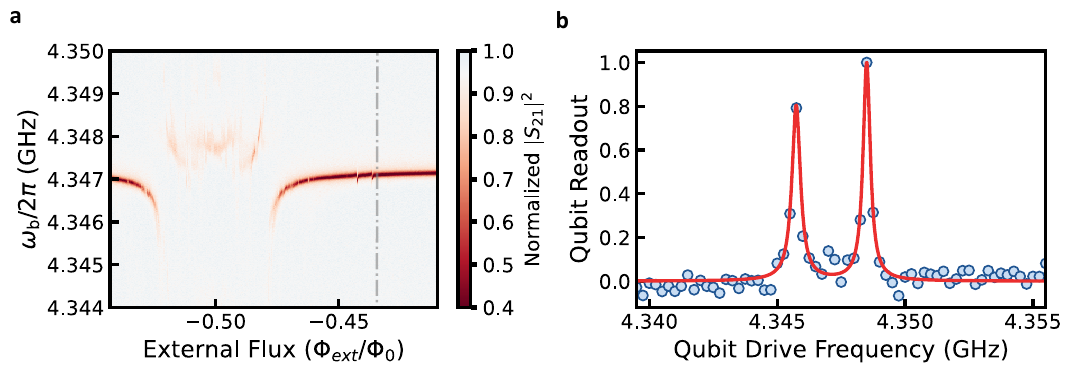}
    \caption{\textbf{Avoided level crossing between the qubit and linear resonator} \textbf{a}, Measured transmission spectra $|S_{21}|^2$ of the linear circuit as the qubit --- shown by the grey dashed line --- is swept through the linear resonator's frequency. Additional avoided crossings are observed as the qubit tunes to its lower sweet spot; the origin of these additional splittings near a half flux quantum is currently being investigated. \textbf{b}, Two-tone spectroscopy of the qubit fully hybridized with the linear resonator. We can extract the linear coupling coefficient $g_{\rm ab}$ between the linear circuit and the qubit from the avoided crossing.}
    \label{SIFig_Avoided}
\end{figure*}

Flux tuning the qubit to be approximately resonant with the linear resonator results in an avoided level crossing, as shown in Fig.~\ref{SIFig_Avoided}. The dipole coupling is given by the frequency difference between the upper and lower normal modes when the qubit and linear circuit are resonant. We extract a dipole coupling rate of $g_{\rm ab} \approx 2 \pi \times 1.4$ MHz, several orders of magnitude smaller than current state-of-the-art circuit quantum electrodynamics devices \cite{blais2021circuit}. The qubit operational point used in this work was detuned from the linear circuit by $\Delta = 1.75$ GHz; therefore, we estimate a critical photon number within the linear circuit of $n_{\rm crit} \approx 350\,000$. 

\subsection{Longitudinal Coupling}

Longitudinal coupling arises from a radiation-pressure-like interaction between the qubit and linear resonator. This coupling modulates the qubit frequency due to the mutual inductive coupling. We can consider the modulation of the qubit frequency by Taylor expanding about a set flux point, given by
\begin{equation}
\label{freqeqn}
    \omega_{\rm q}(\Phi) = \omega_{\rm q,0} + \frac{\partial \omega_{\rm q}}{\partial \Phi}\Phi_{\rm zpf}(\hat{b}+ \hat{b}^{\dagger}) + \frac{1}{2}\frac{\partial^2 \omega_{\rm q}}{\partial \Phi^2}\Phi_{\rm zpf}^2(\hat{b}+ \hat{b}^{\dagger})^2 + \cdots.
\end{equation}
Where $\hat{b}^{(\dagger)}$ is the creation (annihilation) operator for the linear circuit. Inserting Eq.~\ref{freqeqn} into the Hamiltonian $\hat{\mathcal{H}} = \hbar \omega_{\rm q} \hat{a}^{\dagger}\hat{a}$, we arrive at
\begin{equation}
    \mathcal{\hat{H}}/\hbar = \bigg( \omega_{\rm q,0} + \frac{\partial \omega_{\rm q}}{\partial \Phi}\Phi_{\rm zpf}(\hat{b}+ \hat{b}^{\dagger}) + \frac{1}{2}\frac{\partial^2 \omega_{\rm q}}{\partial \Phi^2}\Phi_{\rm zpf}^2(\hat{b}+ \hat{b}^{\dagger})^2 + \cdots \bigg) \hat{a}^{\dagger}\hat{a}.
\end{equation}
We can rewrite this in the form
\begin{equation}
    \mathcal{\hat{H}}/\hbar = \omega_{\rm q,0}\hat{a}^{\dagger}\hat{a} + g_0\hat{a}^{\dagger}\hat{a}(\hat{b}+ \hat{b}^{\dagger}) + g_{\rm q}\hat{a}^{\dagger}\hat{a}(\hat{b}+ \hat{b}^{\dagger})^2 + \cdots.
\end{equation}
Where we can recognize the second term as the longitudinal interaction we are interested in and the third term as the quadratic longitudinal interaction. 

\subsection{Coupling Rate Estimation}

The mutual inductive coupling between the inductor wire of the linear circuit and the SQUID loop of the qubit quantifies the coupling between the linear circuit and the qubit. As shown in Fig.~\ref{SIFig_Image}, the SQUID loop of the qubit has a designed length $l = 120 \mu$m and width $w = 3 \mu$m, giving a nominal SQUID area of $360 \mu$m$^2$; however, the final dimensions were slightly larger due to over-etching. The $1.0 \mu$m wide inductor wire passes $\sim 1.0 \mu$m from the SQUID loop, with measured geometric distances $d_{\rm 1} \approx 1.8 \mu$m and $d_{\rm 2} \approx 5.3 \mu$m are the smallest and largest distance between the inductor and the SQUID loop, respectively; see Fig.~\ref{SIFig_Image}. The mutual inductance can be calculated by integrating the magnetic flux generated by the inductor wire through the SQUID loop and is given by
\begin{equation}
    M = \frac{\mu_0 l}{2\pi}\ln\bigg( \frac{d_{\rm 2}}{d_{\rm 1}}\bigg).
\end{equation}
Using this approximation, we find a mutual inductance $M \approx 25.9$ pH. Moreover, we can determine the zero-point fluctuations of the magnetic flux using the zero-point current, giving $\Phi_{\rm zpf} = MI_{\rm zpf} \approx 461 \: \mu\Phi_0$ where we have written the zero-point magnetic-flux fluctuations in units of flux quanta $\Phi_0 = h/(2e)$.

The longitudinal coupling rate is defined as
\begin{equation}
    g_0 = \frac{\partial\omega_{\rm q}}{\partial\Phi}\Phi_{\rm zpf}.
\end{equation}
The longitudinal coupling rate is the frequency shift of the qubit due to the zero-point fluctuations of the magnetic flux generated by the linear circuit. The frequency responsivity of the qubit can be determined by sweeping the external magnetic field at the operation frequency; we find $\vert \partial\omega_{\rm q} / \partial \Phi \vert \approx 2\pi \times26.0$ GHz/$\Phi_0$. Therefore, at the operation point, we estimate a longitudinal coupling rate of $g_0 \approx 2\pi \times 12.0$ MHz, which is in excellent agreement with the experimentally measured value. 

\begin{figure*}[t]
    \centering
    \includegraphics[width = 0.9\textwidth]{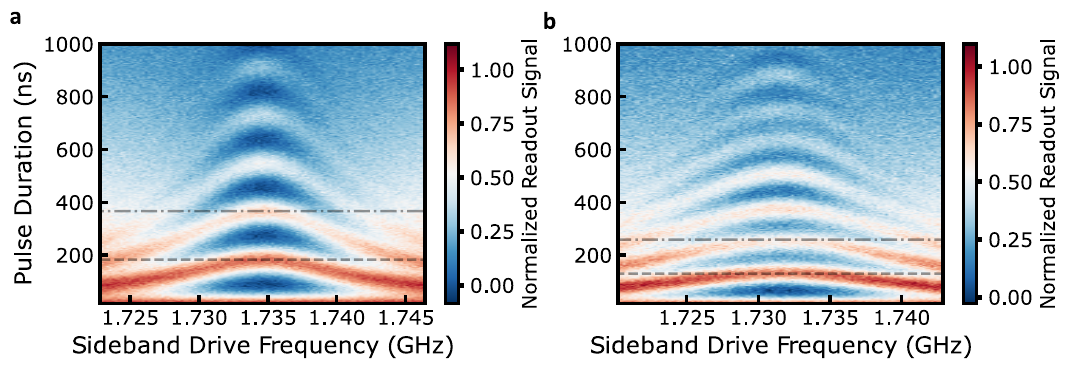}
    \caption{\textbf{On-demand Jaynes-Cummings Ladder} \textbf{a}, Rabi chevrons with a single excitation in the system. A Rabi frequency $g = 2\pi \times 2.73$ MHz was measured. \textbf{b}, Rabi chevrons with two total excitations in the system. The linear resonator is prepared into the Fock state $\vert 1 \rangle$ using a $\pi$-pulse. Following the state preparation, the Rabi chevron was measured. An increased Rabi frequency $g_2 \approx \sqrt{2} g$. The dashed lines indicate the Rabi oscillation period, and the dotted dashed lines indicate two Rabi oscillation periods.}
    \label{SIFig_Rabi2}
\end{figure*}

\subsection{Quadratic Coupling}

If we consider the second-order Taylor expansion of the coupling, we can define the quadratic coupling between the qubit and linear resonator as
\begin{equation}
    \mathcal{\hat{H}}_{\rm q}/\hbar = g_{\rm q}\hat{a}^{\dagger}\hat{a}(\hat{b}+ \hat{b}^{\dagger})^2,
\end{equation}
Where the quadratic coupling rate is defined as,
\begin{equation}
    g_{\rm q} = \frac{1}{2}\frac{\partial^2\omega_{\rm q}}{\partial\Phi^2}\Phi_{\rm zpf}^2.
\end{equation}
Interestingly, the quadratic coupling is non-zero at the qubit sweet spot, where the linear coupling is precisely equal to zero. From the flux arc, at the sweet spot and the qubit operation point, we can estimate the quadratic coupling rate as $g_{\rm q} \approx 2\pi \times5.5$ kHz, and $g_{\rm q} \approx 2\pi \times14.0$ kHz, respectively. Keeping only the non-rotating terms, this quadratic coupling results in a cross-Kerr style coupling of the form,
\begin{equation}
    \mathcal{\hat{H}}_{\rm q}/\hbar \rightarrow \chi_{\rm q}\hat{a}^{\dagger}\hat{a}\hat{b}^{\dagger}\hat{b}.
\end{equation}
Where the value of the cross-Kerr coupling is $\chi_{\rm q} = 2g_{\rm q} \approx 28.0$ kHz.

It should be noted that the bare-to-dressed transition discussed in Section~\ref{paracoup} is a result of the breakdown of a dispersive approximation. The non-perturbative cross-Kerr resulting from the quadratic interaction should not result in a critical photon number \cite{chapple2024robustness}. However, the non-perturbative cross-Kerr coupling may play a role in qubit dephasing. 

\subsection{On-Demand Jaynes-Cummings}

To confirm that the interaction is indeed described by the Jaynes-Cummings Hamiltonian, we compared the single and two-excitation Rabi frequencies. The single excitation Rabi frequency was measured as described in the main text. The data was fit using a numerical simulation described below, where we excite the qubit into the state $\vert 1 \rangle$ using a resonant $\pi$-pulse. To determine the two-excitation Rabi frequency, we first initialize the linear resonator into the Fock state $\vert 1 \rangle$ by performing a sideband $\pi$ pulse. This was followed by a measurement of the Rabi chevron, shown in Fig.~\ref{SIFig_Rabi2}(b). We observe a clear increase in the Rabi frequency as predicted by the Jaynes-Cummings model; the Rabi frequency increases by a factor of approximately square root two. We use our numerical simulation to determine the Rabi frequency given by $g = 2\pi \times 2.76$ MHz, which is in good agreement with a fit of the line-cut to a decaying sinusoidal function giving a value $g = 2\pi \times 2.73$ MHz. We attribute any small discrepancy to the fidelity of our initial preparation of the linear resonator Fock state. Due to losses within the system, on average, the state of the linear resonator is likely a mixed state of $\vert 0 \rangle$ and $\vert 1 \rangle$, reducing the ensemble-averaged Rabi frequency measured here. However, this measurement does confirm that the interaction produces a Jaynes-Cummings ladder. With device optimization, this will be explored in future work. 

\subsection{Master Equation Simulation}

To model the dynamics of the drive-induced Jaynes-Cummings interaction, we simulate the system using the Lindblad Master equation using the Python package Qutip \cite{johansson2012qutip}. The dynamics are governed by the equation given by,
\begin{equation}
    \frac{\partial \rho}{\partial t} = -\frac{i}{\hbar}[ \mathcal{\hat{H}}(t),\rho ] + \kappa_{\rm{b}} \mathcal{L}[\hat{b}] \rho + \gamma_1 \mathcal{L}[\hat{\sigma}_-] \rho,
\end{equation}
where $\rho$ is the system's density matrix, $\kappa_{\rm b}$ is the total decay rate of the linear microwave resonator, and $\gamma_1$ is the qubit relaxation rate. We do not consider dephasing here since single-shot projections on the z-axis are insensitive to fast frequency fluctuations, and $T_{2 \rm echo}$ in our experiment is much greater than $T_1$ \cite{abad2022universal}. The system Hamiltonian derived below is given by, 
\begin{equation}
    \mathcal{\hat{H}}(t)/\hbar = -\frac{\Delta_{\rm q}}{2} \hat{\sigma}_{\rm z} + \omega_{\rm b}\hat{b}^{\dagger}\hat{b} + g(t)(\hat{\sigma}_+\hat{b}+\hat{\sigma}_-\hat{b}^{\dagger}) + \epsilon_{\rm d}(t)(\hat{\sigma}_+e^{-i\Delta_{\rm q}t} + \hat{\sigma}_-e^{i\Delta_{\rm q}t}).
\end{equation}
Here, $\Delta_{\rm q} = \omega_{\rm d} - \omega_{\rm q}$ is the drive detuning, $g(t)$ is the time dependent coupling rate, and $\epsilon_{\rm d}(t)$ is a time dependent resonant qubit drive. The system is initially prepared in its ground state, and the qubit is prepared with a sequence of resonant square wave $\pi$-pulses $24$ ns in duration and sideband $\pi$-swap pulses $91$ ns in duration. The resultant time traces are plotted in Fig.~3 in the main text and show excellent agreement with the experimental results. 

\section{Longitudinal Coupling Theory}

\subsection{Quantization of Superconducting Qubit}

As defined in Ref.~\cite{Krantz_Quantum_2019}, the Hamiltonian describing an LC oscillator is given by
\begin{equation}
    \mathcal{\hat{H}} = 4 E_{\rm C}\hat{n}^2 + \frac{1}{2}E_{\rm L}\hat{\phi}^2. 
\end{equation}
Here, $E_{\rm C} = e^2/(2C)$, is the charging energy required to add each electron from the Cooper-pair to the island and $E_{\rm L} =  (\Phi_{\rm 0}/2\pi)^2 /L$ is the inductive energy, where $\Phi_{\rm 0} = h/(2e)$ is the magnetic flux quantum. The operator $\hat{n} =  \hat{Q}/2e$ is the excess number of Cooper pairs on the island, and $\hat{\phi}$ is the gauge invariant phase across the inductor. Note that these form a pair of canonically conjugate variables $[\hat{\phi},\hat{n}]=i$.

We can recast this Hamiltonian in the form,
\begin{equation}
    \mathcal{\hat{H}} = \hbar\omega_{\rm q}\big( \hat{a}^{\dagger}\hat{a}+1/2\big),
\end{equation}
where $\hat{a} (\hat{a}^{\dagger})$ is the annihilation (creation) operator of a single excitation of the LC oscillator. The resonance frequency is defined as $\omega_{\rm q} = \sqrt{8E_{\rm C}E_{\rm L}}/\hbar$ in anticipation of this circuit representing the transmon qubit. The charge and flux operators can be written in the form $\hat{n} = in_{\rm zpf} (\hat{a}-\hat{a}^{\dagger})$ and $\hat{\phi} = \phi_{\rm zpf}(\hat{a}+\hat{a}^{\dagger})$, where $n_{\rm zpf} = (E_{\rm L}/ (32E_{\rm C}))^{1/4}$ and $\phi_{\rm zpf} = (2E_{\rm C}/E_{\rm L})^{1/4}$.

Replacing the inductor with a Josephson junction results in the modified Hamiltonian of the form,
\begin{equation}
    \mathcal{\hat{H}} = 4 E_{\rm C}\hat{n}^2 - E_{\rm J}\cos{(\hat{\phi})},
\end{equation}
where $E_{\rm C} = e^2 / (2C_{\Sigma})$, $C_{\Sigma} = C_{\rm s} + C_{\rm J}$ is the total capacitance, including the shunt capacitance and the self-capacitance of the Josephson junction, $E_{\rm J} = I_{\rm c} \Phi_{0}/2\pi$ is the Josephson energy, where $I_{\rm c}$ is the critical current of the junction. In the transmon limit (i.e. $E_{\rm J} \gg E_{\rm C}$), the fluctuations of $\hat{\phi}$ are small; therefore, we can expand the potential term, giving
\begin{equation}
    -E_{\rm J}\cos{(\hat{\phi})} = \frac{1}{2}E_{\rm J}\hat{\phi}^2-\frac{1}{24}E_{\rm J}\hat{\phi}^4 + \mathcal{O}(\hat{\phi}^6).
    \label{EJ_Expan}
\end{equation}
The second term results in an anharmonic potential, with an anharmonicity given by $\alpha = \omega_{\rm q}^{1\rightarrow2}-\omega_{\rm q}^{0\rightarrow1}$. For a transmon qubit $\alpha = -E_{\rm C}$, it is typically 100-300 MHz. 

Keeping terms up to fourth order in creation and annihilation operators, the Hamiltonian can be written in the form,
\begin{equation}
    \mathcal{\hat{H}}/\hbar = \omega_{\rm q}\hat{a}^{\dagger}\hat{a} + \frac{\alpha}{2}\hat{a}^{\dagger}\hat{a}^{\dagger}\hat{a}\hat{a},
\end{equation}
which is the Hamiltonian for a Kerr oscillator. Rather than a single Josephson junction, a pair of junctions are used as a SQUID loop to provide flux tunability to the transmon qubit. The effective Hamiltonian of such a flux-tunable transmon is given by
\begin{equation}
    \mathcal{\hat{H}} = 4 E_{\rm C}\hat{n}^2 -2E_{\rm J}\vert \cos{(\varphi_{\rm e})} \vert \cos{\hat{\phi}}. 
\end{equation}
Here $\varphi_{\rm e}$ is the externally applied magnetic flux. We can define an effective Josephson energy $\Tilde{E}_{\rm J}(\varphi_{\rm e}) = 2E_{\rm J}\vert \cos{(\varphi_{\rm e})} \vert$.

\subsection{Longitudinal Hamiltonian}

Considering the linear resonator and the coherent drive tones, the total Hamiltonian of the device is
\begin{equation}
    \mathcal{\hat{H}} = 4 E_{\rm C}\hat{n}^2 - \Tilde{E}_{\rm J}(\varphi_{\rm e})\cos{\hat{\phi}} + \hbar \omega_{\rm b} \hat{b}^{\dagger}\hat{b} + \mathcal{\hat{H}}_{\rm d}.
\end{equation}
Here, $\hat{b} (\hat{b}^{\dagger})$ are the annihilation (creation) operators for the linear resonator. During operation, the external flux bias is fixed at a single value, and due to the mutual inductive coupling between the linear circuit and the SQUID loop, the external flux is modulated by the linear resonator. We can, therefore, write the Josephson energy in the form
\begin{equation}
     \tilde{E}_{\rm J}(\varphi_{\rm e}) = \bar{E}_{\rm J} + \frac{\mathrm{d} \bar{E}_{\rm J}}{\mathrm{d} \varphi_{\rm e}}\frac{\mathrm{d} \varphi_{\rm e}}{\mathrm{d} I} I_{zpf}(\hat{b} + \hat{b}^{\dagger}).
\end{equation}
Here, $\Bar{E}_{\rm J}$ is the Josephson energy at the flux-bias point, $I_{\rm zpf} = \sqrt{\hbar \omega_{\rm b} / 2L_{\rm b}}$, where $L_{\rm b}$ is the inductance of the linear resonator. We can simplify this equation by defining a coupling constant in the form
\begin{equation}
    \hbar \Tilde{g}_{0} = \frac{\mathrm{d} \bar{E}_{\rm J}}{\mathrm{d} \varphi_{\rm e}}\frac{\mathrm{d} \varphi_{\rm e}}{\mathrm{d} I} I_{zpf}.
\end{equation}
Therefore, the Hamiltonian can be written as,
\begin{equation}
    \mathcal{\hat{H}} = 4 E_{\rm C}\hat{n}^2 - \Bar{E}_{\rm J}\cos{\hat{\phi}} - \hbar \Tilde{g}_0\cos{\hat{\phi}}(\hat{b} + \hat{b}^{\dagger}) + \hbar \omega_{\rm b} \hat{b}^{\dagger}\hat{b} + \mathcal{\hat{H}}_{\rm d}.
\end{equation}
At this stage, we can expand the cosine of the gauge invariant flux using Eq.~\ref{EJ_Expan}, giving
\begin{equation}
    \mathcal{\hat{H}} = 4 E_{\rm C}\hat{n}^2 + \Bar{E}_{\rm J}(\frac{1}{2}\hat{\phi}^2-\frac{1}{24}\hat{\phi}^4 ) + \hbar \Tilde{g}_0\frac{1}{2}\hat{\phi}^2(\hat{b} + \hat{b}^{\dagger}) + \hbar \omega_{\rm b} \hat{b}^{\dagger}\hat{b} + \mathcal{\hat{H}}_{\rm d}.
    \label{Exact_Ham}
\end{equation}
Keeping only first-order terms in the interaction and using the definitions $\hat{\phi} = \phi_{\rm zpf}(\hat{a}+\hat{a}^{\dagger})$ and $\phi_{\rm zpf} = (2E_{\rm C}/\bar{E}_{\rm J})^{1/4}$ we arrive at the Hamiltonian based on creation and annihilation operators. The Hamiltonian can be written in the form,
\begin{equation}
    \mathcal{\hat{H}}/\hbar = \omega_{\rm q} \hat{a}^{\dagger}\hat{a} + \frac{\alpha}{2}\hat{a}^{\dagger}\hat{a}^{\dagger}\hat{a}\hat{a} + \omega_{\rm b} \hat{b}^{\dagger}\hat{b} + \frac{\Tilde{g}_0 }{2}\bigg( \frac{2E_{\rm C}}{\bar{E}_{\rm J}} \bigg)^{1/2} ( \hat{a} + \hat{a}^{\dagger})^2 (\hat{b} + \hat{b}^{\dagger}) + \mathcal{\hat{H}}_{\rm d}/\hbar.
    \label{Exact_Ham}
\end{equation}
Where, we will define $g_0 = \Tilde{g}_0(2E_{\rm C}/\bar{E}_{\rm J})^{1/2}$. Moreover, the drive Hamiltonian can be written in the form,
\begin{equation}
    \mathcal{\hat{H}}_{\rm d} =  \hbar\epsilon_{\rm d} (\hat{a}e^{i\omega_{\rm d}t}  + \hat{a}^{\dagger}e^{-i\omega_{\rm d}t} )
\end{equation}
Here, $\epsilon_{\rm d} = \sqrt{\mathcal{P}\kappa_{\rm e}/\hbar \omega_{\rm d}}$ is the coherent drive strength. We can now transform the Hamiltonian into a frame rotating at the drive frequency using the unitary transformation $\hat{\mathcal{U}} = e^{i\omega_{\rm d} \hat{a}^{\dagger}\hat{a} t}$, where $\mathcal{\hat{H}}_{\rm rot} = \hat{\mathcal{U}}\mathcal{\hat{H}}\hat{\mathcal{U}}^{\dagger} + i\hbar (\mathrm{d}\hat{\mathcal{U}}/\mathrm{d}t)\hat{\mathcal{U}}^{\dagger}$, keeping only co-rotating terms we can express the new Hamiltonian as
\begin{equation}
    \mathcal{\hat{H}}/\hbar = -\Delta_{\rm q} \hat{a}^{\dagger}\hat{a} + \frac{\alpha}{2}\hat{a}^{\dagger}\hat{a}^{\dagger}\hat{a}\hat{a} + \omega_{\rm b} \hat{b}^{\dagger}\hat{b} + g_0\hat{a}^{\dagger}\hat{a}(\hat{b} + \hat{b}^{\dagger}) + \epsilon_{\rm d}(\hat{a} + \hat{a}^{\dagger}).
    \label{PP_Ham}
\end{equation}
Here $\Delta_{\rm q} = \omega_{\rm d} - \omega_{\rm q}$. We can first study the steady-state classical dynamics of the Kerr oscillator. The classical equation of motion is given by
\begin{equation}
    \frac{\mathrm{d}}{\mathrm{d}t}\langle \hat{a} \rangle = (i\Delta_{\rm q} - \gamma_{\rm q}/2)\langle \hat{a} \rangle - i \alpha \vert \langle \hat{a} \rangle \vert^2 \langle \hat{a} \rangle - i g_0 (\langle \hat{b} \rangle + \langle \hat{b}\rangle^*)\langle \hat{a} \rangle - \epsilon_{\rm d}.
\end{equation}
The coupling to the linear resonator mode has the effect of modifying the Kerr nonlinearity as described in \cite{Zoepfl_Kerr_2023}, with an effective Kerr nonlinearity of
\begin{equation}
    \alpha_{\rm eff} = \alpha - \frac{2g_0^2\omega_{\rm b}}{\omega_{\rm b}^2 + \gamma_{\rm b}^2/4}.
\end{equation}
Where $\gamma_{\rm b}$ is the linewidth of the linear mode. Using the parameters from our device, the correction to the Kerr nonlinearity is much less than the Kerr nonlinearity. Therefore, the equation of motion can be written in the form,
\begin{equation}
    \frac{\mathrm{d}}{\mathrm{d}t}\langle \hat{a} \rangle = (i\Delta_{\rm q} - \gamma_{\rm q}/2)\langle \hat{a} \rangle - i \alpha \vert \langle \hat{a} \rangle \vert^2 \langle \hat{a} \rangle - \epsilon_{\rm d}.
\end{equation}
In steady-state (i.e. $\langle \Dot{\hat{a}}\rangle = 0$), the cubic equation has one real root, but for a large enough input drive, a bistable regime exists. The cubic equation has the form,
\begin{equation}
    \Bar{n}_{\rm q}(-\Delta_{\rm q} + \alpha \Bar{n}_{\rm q})^2 - \epsilon_{\rm d}^2 = 0.
    \label{Cubic_eqn}
\end{equation}
Here, we have defined $\Bar{n}_{\rm q} = \vert \langle \hat{a} \rangle \vert^2$ the steady-state amplitude of the Kerr oscillator.

\subsection{Photon Number Calibration}

To calibrate the number of photons in the qubit due to the sideband drive tone, we use the AC-Stark shift. The driving term applied on the red-sideband results in an AC-Stark shift of the Kerr oscillator. Since the drive tone detuning is large ($\Delta_{\rm q} = \omega_{\rm d} - \omega_{\rm q} \approx \omega_{\rm b} = 4.35$ GHz), we must consider the Bloch-Siegert shift due to the counter-rotating terms, which is described in Ref.~\cite{Ann_Sideband_2021}. The AC-Stark shift considering the co- and counter-rotating terms is given by,
\begin{equation}
    \delta\omega_{\rm q} = \frac{1}{2}\epsilon_{\rm d}^2 \alpha \bigg( \frac{1}{\Delta_{\rm q}^2} + \frac{2}{\vert \Delta_{\rm q}\vert \Sigma} + \frac{1}{\Sigma^2} \bigg).
\end{equation}
Where $\Sigma = \omega_{\rm q} + \omega_{\rm d}$, $\alpha$ is the transmon anharmonicity and $\epsilon_{\rm d} = \sqrt{\mathcal{P}\kappa_{\rm e}/\hbar \omega_{\rm d}}$. We can re-arrange this equation by writing it in the form,
\begin{equation}
    \epsilon_{\rm d}^2 = \frac{2 \delta\omega_{\rm q}}{\alpha}\bigg( \frac{1}{\Delta_{\rm q}^2} + \frac{2}{\vert \Delta_{\rm q}\vert \Sigma} + \frac{1}{\Sigma^2} \bigg)^{-1}.
    \label{Stark_eqn}
\end{equation}

Therefore, we can calibrate the average photon number in the Kerr oscillator using the AC-Stark shift. Combining Eq.~\ref{Cubic_eqn} and Eq.~\ref{Stark_eqn}, the average photon number can be found by solving the cubic equation,
\begin{equation}
    \Bar{n}_{\rm q}(-\Delta_{\rm q} + \alpha \Bar{n}_{\rm q})^2 - \frac{2 \delta\omega_{\rm q}}{\alpha}\bigg( \frac{1}{\Delta_{\rm q}^2} + \frac{2}{\vert \Delta_{\rm q}\vert \Sigma} + \frac{1}{\Sigma^2} \bigg)^{-1}= 0.
    \label{n_bar_eq}
\end{equation}

The first real solution to this equation is the photon population of the Kerr-oscillator, $\langle \hat{a}^{\dagger}\hat{a} \rangle = \Bar{n}_{\rm q}$. Note that treating the qubit as a two-level system at this step underestimates the sideband population and, therefore, will overestimate $g_0$. 

\section{Linearized Coupling}

At this point, to continue our analysis, we aim to linearize the theory about the sideband drive applied to the qubit. From the AC-Stark shift calibration above, we found that for all drive powers, the average occupancy $\vert \langle \hat{a} \rangle \vert^2$ is much less than a single photon on average. Therefore, we will perform our analysis assuming a harmonic oscillator, making the two-level approximation once we have arrived at a linearized theory. 

\subsection{Sideband Linearization}

If we begin with the Hamiltonian given by Eq.~\ref{Exact_Ham}, rewriting it in the form,
\begin{equation}
        \mathcal{\hat{H}}/\hbar = \omega_{\rm q} \hat{a}^{\dagger}\hat{a} + \frac{\alpha}{2}\hat{a}^{\dagger}\hat{a}^{\dagger}\hat{a}\hat{a} + \omega_{\rm b} \hat{b}^{\dagger}\hat{b} + \frac{g_0}{2}( \hat{a}^2 + (\hat{a}^{\dagger})^2 + 2\hat{a}^{\dagger}\hat{a}+1) (\hat{b} + \hat{b}^{\dagger}) + (\epsilon_{\rm d}^*\hat{a}e^{i\omega_{\rm d}t}  + \epsilon_{\rm d}\hat{a}^{\dagger}e^{-i\omega_{\rm d}t} ).
        \label{Lin_NonLin_PP}
\end{equation}
In the experimentally relevant situation where the sideband drive is always weak such that $\vert \langle \hat{a} \rangle \vert^2 \ll 1$, we will drop the Kerr term as the Kerr oscillator is well approximated by a harmonic oscillator in this weak driving regime. This results in the Hamiltonian 
\begin{equation}
        \mathcal{\hat{H}}/\hbar = \omega_{\rm q} \hat{a}^{\dagger}\hat{a} + \omega_{\rm b} \hat{b}^{\dagger}\hat{b} + g_0\hat{a}^{\dagger}\hat{a}(\hat{b} + \hat{b}^{\dagger}) + (\epsilon_{\rm d}^*\hat{a}e^{i\omega_{\rm d}t}  + \epsilon_{\rm d}\hat{a}^{\dagger}e^{-i\omega_{\rm d}t} ).
\end{equation}
We can simultaneously transform into a frame rotating at the drive frequency and displaced by the steady-state amplitude $\langle \hat{a} \rangle$. This is equivalent to applying the unitary $\hat{\mathcal{D}}\hat{\mathcal{U}} = e^{\langle \hat{a} \rangle^*\hat{a}-\langle \hat{a} \rangle\hat{a}^{\dagger}}e^{i\omega_{\rm d}\hat{a}^{\dagger}\hat{a}t}$. The resulting Hamiltonian in the transformed frame is
\begin{equation}
        \mathcal{\hat{H}}/\hbar = -{\Delta_{\rm q}}\hat{a}^{\dagger}\hat{a} + \omega_{\rm b} \hat{b}^{\dagger}\hat{b} + g_0(\langle \hat{a} \rangle^* \hat{a} + \langle \hat{a} \rangle \hat{a}^{\dagger})(\hat{b} + \hat{b}^{\dagger}) + g_0\hat{a}^{\dagger}\hat{a}(\hat{b} + \hat{b}^{\dagger}). 
\end{equation}
This transformation has separated the interaction into its linear and non-linear components. In this way, we have \textit{only} performed unitary transformations resulting in static shifts of the qubit's equilibrium frequency. Moreover, the last term in this Hamiltonian results in a non-linear two-photon interaction, as described in Ref.~\cite{hauer2023nonlinear}. This term can be dropped since it does not affect the dynamics in the frame of our sideband drive. However, this term is crucial for generating Schrodinger cat states via the non-linearity of the longitudinal interaction. We can write our Hamiltonian in the simplified form,
\begin{equation}
    \mathcal{\hat{H}}/\hbar = -\Delta_{\rm q} \hat{a}^{\dagger}\hat{a} + \omega_{\rm b}\hat{b}^{\dagger}\hat{b} + g(\hat{a}+\hat{a}^{\dagger})(\hat{b}+\hat{b}^{\dagger}).
\end{equation}
We define the enhanced coupling rate as $g = \vert \langle \hat{a} \rangle \vert g_0$ and have assumed it is real. Performing the RWA and approximating the harmonic oscillator as a two-level system, we can write the final Hamiltonian as,
\begin{equation}
    \mathcal{\hat{H}}/\hbar = -\frac{\Delta_{\rm q}}{2} \hat{\sigma}_{\rm z} + \omega_{\rm b}\hat{b}^{\dagger}\hat{b} + g(\hat{\sigma}_+\hat{b}+\hat{\sigma}_-\hat{b}^{\dagger}).
    \label{PP_JC_Ham}
\end{equation}
We arrive at the drive-induced Jaynes-Cummings Hamiltonian. Following standard quantum optics techniques \cite{reiserer2015cavity}, one can derive the $N$ excitation Rabi frequency given by $2g_{N} = 2\sqrt{N}g$. 

\subsection{Normal-Mode Spectrum}

Considering the Hamiltonian given in Eq.~\ref{PP_JC_Ham}, we can derive the eigenfrequencies to fit the normal-mode spectra. If we consider the quantity $\hat{N} = \hat{a}^{\dagger}\hat{a} + \hat{\sigma}_+\hat{\sigma}_-$ which defines the total number of excitations within the system, in the absence of decay. We notice that $[\hat{\mathcal{H}},\hat{N}]=0$, thus is a conserved quantity of the system. Therefore, we can consider the block-diagonal element of the Hamiltonian defined by the total number of excitations spanning the subspace $\{ \vert g, n \rangle, \vert e, n-1 \rangle \}$. This block diagonal element can be written in the form,
\begin{equation}
    \mathcal{\hat{H}}/\hbar = \begin{pmatrix}
    -{\Delta}_{\rm qb} + n\omega_{\rm b} & g\sqrt{n} \\
    g\sqrt{n} & n\omega_{\rm b} 
    \end{pmatrix}.    
\end{equation}
Where $\Delta_{\rm qb} = \omega_{\rm b}+\Delta_{\rm q}$, recalling that in the rotating frame $\Delta_{\rm q} = \omega_{\rm d} - \omega_{\rm q} \approx -\omega_{\rm b}$; therefore, this is the qubit-linear resonator detuning in the rotating frame. Following some algebra, it can be shown that the energy eigenvalues of this block diagonal Hamiltonian have the form \cite{gardiner2004quantum},
\begin{equation}
    E_{\rm n,\pm} = n\hbar\omega_{\rm b} - \hbar\frac{\Delta_{\rm qb}}{2} \pm \frac{\hbar}{2}\sqrt{\Delta_{\rm qb}^2 + 4ng^2},
\end{equation}
from which we can recover the $n$-quanta Rabi frequency $2g_{\rm n} = 2\sqrt{n}g$. If we consider the lowest energy manifold of the Jaynes-Cummings ladder, we can determine the expected spectroscopic eigenfrequencies. For $n=1$, we find the frequency spectrum given by,
\begin{equation}
    \omega_{\pm} = \omega_b - \frac{\Delta_{\rm qb}}{2} \pm \frac{1}{2}\sqrt{\Delta_{\rm qb}^2 + 4g^2}.
\end{equation}
This can be written in the more enlightening form,
\begin{equation}
    \omega_{\pm} = \frac{\omega_{\rm b} -\Delta_{\rm q}}{2} \pm \frac{1}{2}\sqrt{\Delta_{\rm qb}^2 + 4g^2}.
    \label{Norm_Mode}
\end{equation}
Using Eq.~\ref{Norm_Mode}, we can fit the experimental data to extract $g$ as a function of sideband drive power, as shown in the main text Fig.~2. Using the extracted value of $g$ and the photon number extracted from Eq.~\ref{n_bar_eq}, we can determine the single-photon coupling rate $g_0$, as shown in the main text. It is important to note that this solution \textit{only} holds when the system is within the lowest energy manifold of the Jaynes-Cummings ladder. 

\clearpage

\end{document}